\RequirePackage{fix-cm}
\documentclass{article}
\usepackage{hyperref}
\usepackage[left=2.0cm,right=2.0cm,top=2.5cm,bottom=2.5cm]{geometry}
\usepackage{color}
\usepackage{graphicx}
\usepackage{amsmath,amssymb}
\usepackage{caption}
\usepackage{subcaption}

\title{Phase-preserving nonreciprocal dynamics in coupled nonlinear oscillators}

\author{Ali Kogani\footnotemark[1] 
\and Behrooz Yousefzadeh\footnotemark[2]}

\begin{document}
\renewcommand{\thefootnote}{\fnsymbol{footnote}}
\footnotetext[1]{Department of Mechanical, Industrial \& Aerospace Engineering, Concordia University, Montreal, QC, H3G1M8, Canada

  (\href{mailto:ali.kogani@mail.concordia.ca}{ali.kogani@mail.concordia.ca})
}
\footnotetext[2]{Department of Mechanical, Industrial \& Aerospace Engineering, Concordia University, Montreal, QC, H3G1M8, Canada

  (\href{mailto:behrooz.yousefzadeh@concordia.ca}{behrooz.yousefzadeh@concordia.ca})}

\maketitle

\begin{abstract}

Nonreciprocity is most commonly associated with a large difference in the transmitted energy when the locations of the source and receiver are interchanged. This energy bias is accompanied by a difference in the transmitted phase. We highlight the role of this phase bias in breaking reciprocity in the steady-state vibration transmission characteristics of coupled nonlinear systems to external harmonic excitation. We show that breaking of reciprocity is most commonly accompanied by a simultaneous bias in the transmitted energy and phase. Energy bias alone, without any contribution from phase, can still lead to nonreciprocity, but only at very finely tuned system parameters. We provide a methodology for realizing response regimes of phase-preserving nonreciprocity using two independent symmetry-breaking parameters in the system. Our findings highlight the key contribution of phase in nonlinear nonreciprocity.  
\end{abstract}



\maketitle

\section{\label{intro} Introduction}

Reciprocity refers to the symmetry property of wave and vibration phenomena that guarantees transmission characteristics between two points do not depend on the direction of travel. This concept has been extensively studied and applied since the nineteenth century, with seminal contributions from Helmholtz~\cite{von1896theorie} and Rayleigh~\cite{strutt_general_1871}, among many others. Beyond its significant theoretical implications~\cite{achenbach_reciprocity_2003}, reciprocity has underpinned diverse experimental methodologies in fields such as vibroacoustics~\cite{fahy_applications_2003,ten_wolde_reciprocity_2010}, structural dynamics~\cite{ewins_modal_2009,TPA}, defect detection, determination of elastic constants~\cite{auld_general_1979}, ultrasonics~\cite{anderson_use_1998}, and seismology~\cite{knopoff_seismic_1959}. 

For a device to allow for different transmission properties in opposite directions, it needs to operate beyond the bounds of reciprocity. Nonreciprocity plays a crucial role in the functioning of well-established communication devices such as isolators and circulators~\cite{reiskarimian_nonreciprocal_2019}. A celebrated example of nonreciprocity in vibration and acoustics is the mechanical or acoustic diode, which restricts waves to travel in only one direction~\cite{liang2010acoustic, boechler_bifurcation-based_2011,GRINBERG201849}. Nonreciprocity also facilitates wave filtering and frequency conversion. To name a few examples from one-dimensional systems, high-efficiency, broadband acoustic waveguides capable of converting wave frequencies have been proposed for potential applications in sonar and ultrasound imaging~\cite{Fu_freq_conver}, resonators with reconfigurable bandwidth properties have been developed~\cite{tsakmakidis2017breaking}, and nonreciprocity has played an important role in enabling and enhancing energy localization and irreversible energy transmission in mechanical systems~\cite{TET,wang2020irreversible}. 

Nonreciprocal dynamics in mechanical systems can arise through several different mechanisms~\cite{nassar_nonreciprocity_2020}. One approach ({\it active}) involves time-dependent modulation of the effective properties by means of external controls, such as introducing kinetic motion or applying spatiotemporal modulations~\cite{ZANGENEHNEJAD2019100031, nassar_modulated_2017, wu2025linear}. 
Another approach ({\it passive}) relies on nonlinear forces within the system~\cite{lepri_asymmetric_2011, blanchard_non-reciprocity_2018, moore_nonreciprocity_2018}. Nonlinearity introduces various mechanisms that drive nonreciprocity, including the dependence of the response on the amplitude of motion, generation of higher harmonics, and various bifurcations. Regardless of the approach, a system with mirror symmetry cannot support a nonreciprocal response because symmetry ensures the transmission between the two points is identical in both directions. Thus, breaking the mirror symmetry is a necessary (but not sufficient~\cite{giraldo_restoring_2023}) condition for enabling nonreciprocity.

We focus exclusively on nonreciprocal dynamics in nonlinear systems in this work. Structural asymmetry may be introduced locally within the system in form of a defect~\cite{boechler_bifurcation-based_2011}, periodically throughout the structure~\cite{darabi_broadband_2019, PATIL2022101821}, as an effective gate by combining two mirror-symmetric sub-structures~\cite{gatePai,lu_unilateral_2021}, or by incorporating nonreciprocal internal forces~\cite{brandenbourger_non-reciprocal_2019}. In all cases, the most salient indicator of nonreciprocity is the ability of a system to support unidirectional (diode-type) transmission. This occurs when there is a large difference between the energies transmitted in opposite directions. This prominent feature of nonreciprocity has driven the primary interest in the study of nonreciprocal dynamics. 

In addition to a difference in transmitted energies (energy bias), nonreciprocity is accompanied by a difference in the phase of the transmitted vibrations (phase bias). The phase bias, however, is often overlooked. The extreme case of this phenomenon occurs when the transmitted energies between two points remain unchanged upon interchanging of the source and receiver, but there is still a difference in the transmitted phases. This effect, {\it phase nonreciprocity}, has been shown for the steady-state response to external harmonic excitation, both for a system with two degrees of freedom (coupled waveguides)~\cite{yousefzadeh_computation_2022} and a spatially periodic system~\cite{kogani_nonreciprocal_2024}. The resulting nonreciprocal phase shift is the only contributor to nonreciprocity in this case.

In this work, while still focusing on the role of phase in nonreciprocity, we tackle a different question: do nonreciprocal response regimes exist that are characterized by equal transmitted phases but different transmitted energies? We refer to such response regimes as {\it phase-preserving nonreciprocity}. Energy bias has been the most common indicator of nonreciprocity so far. Phase-preserving nonreciprocity will determine whether an energy bias alone (without contribution from phase) can lead to nonreciprocity. Thereby, this investigation highlights yet another aspect of the contribution of phase in breaking reciprocity. 

Nonreciprocity in the transmitted phase has been a subject of investigation in electronics and optics~\cite{kamal1960parametric,yokoi_demonstration_1999}, with recent applications in optical and acoustic waveguides~\cite{yokoi_calculation_2008, tzuang_non-reciprocal_2014,verba_phase_2021, zangeneh2018doppler}. The ability to passively control the direction-dependent transmitted phase of a waveguide may find application in vibration control strategies or in performing certain logic operations~\cite{wang2021solving}.

We use a lumped-parameter model to investigate phase-preserving nonreciprocity. Lumped-parameter models represent the phenomenon in systems that can be adequately modeled as a combination of scalar wave fields and coupled oscillators. These models have been widely used in acoustics and vibrations to describe wave propagation and resonance phenomena. In phononic crystals and metamaterials, lumped-parameter models are capable of presenting a concise description of complex physics such as formation of Bragg and sub-Bragg bandgaps~\cite{hussein2014dynamics,phani2017elastodynamics}, bandgaps induced by inertial amplification~\cite{yilmaz2007phononic}, directional bandgaps in spatiotemporally modulated systems~\cite{swinteck2015bulk,nassar2017non,hasan2019geometric}, amplitude-dependent bandgaps in nonlinear systems~\cite{vakakis1995nonlinear,leamy2011},  cloaking~\cite{chen2021discrete}, flat bands~\cite{matlack2018designing} and topological effects~\cite{susstrunk2016classification,chen2018study,matlack2018designing}, to name a few examples. The present work is carried out within the same context.

Following the previous work on phase nonreciprocity~\cite{yousefzadeh_computation_2022,kogani_nonreciprocal_2024}, we investigate phase-preserving nonreciprocity in the steady-state response of two nonlinear oscillators to external harmonic excitation. Our methodology for finding phase-preserving nonreciprocity relies on first establishing phase nonreciprocity as an intermediate operating point. 
Here, the phase of the response refers to the angular relationship between the steady-state displacement and the external force, which represents the delay or advance between the input and output of the system over one cycle of oscillation. Phase-preserving nonreciprocity, therefore, refers to the scenario in which this phase shift remains unchanged when the locations of the source and receiver are interchanged.

Section~\ref{sec:intro} introduces the system under investigation and the solution methodology. Section~\ref{sec:control} presents the procedure that enables us to find a family of system parameters that lead to phase-preserving nonreciprocity. This involves finding response regimes that exhibit phase nonreciprocity and reciprocal dynamics. Section~\ref{sec:main} presents the main results on phase-preserving nonreciprocity. We summarize our findings in Section~\ref{sec:conclusion}. 

\section{Problem Setup and Methodology}
\label{sec:intro}

\begin{figure}
  \includegraphics[width=.35\textwidth]{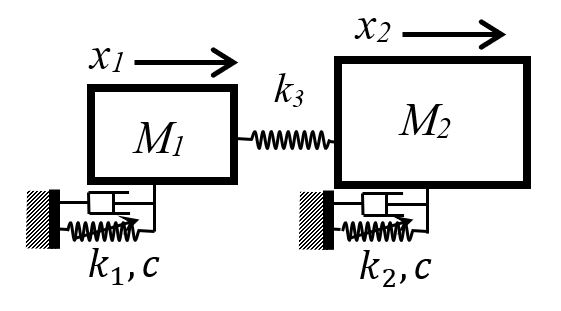}
  \centering
\caption{Schematic representation of the 2DoF system.}
\label{fig:1}      
\end{figure}

Fig.~\ref{fig:1} shows a schematic representation of the two-degree-of-freedom (2DoF) system we study in this work. The system consists of two masses, $M_1$ and $M_2 = \mu M_1$, that are coupled  by a linear spring of constant $k_3$. The mass $M_1$ is anchored to the ground by a spring with cubic nonlinearity, $k_1=k_{g1}+k_{n1}\delta^2$, where $\delta$ represents the spring deformation from its static equilibrium position. The mass $M_2$ is also anchored to the ground with a similar nonlinear spring of constant $k_2=k_{g2}+k_{n2}\delta^2$. Energy dissipation is modeled by a linear viscous damping mechanism, represented by a dashpot of constant $c$ connecting each mass to the ground. The system is subject to an external harmonic force of amplitude $F$ and frequency $\omega_f$ (not shown). 

The mirror symmetry of the system is controlled by the ratio of the two masses, $\mu=M_2/M_1$, the ratio of the two grounding linear springs, $r=k_{g2}/k_{g1}$, or the ratio of the nonlinear spring coefficients, $\alpha=k_{n2}/k_{n1}$. We note that independent tuning of the linear and nonlinear portions of the effective elasticity of the system is already reported in the literature~\cite{Frazier,patil_review_2022}. 

As outlined in Appendix A, the equation of motion of the system can be expressed in non-dimensional parameters as follows:
\begin{equation}
\label{govern}
\begin{aligned}
\ddot{x}_1+k_c(x_1-x_2)+x_1+k_Nx_1^3+2\zeta\dot{x}_1=F_1\cos{\omega_ft} \\
\mu\ddot{x}_2+k_c(x_2-x_1)+rx_2+\alpha k_Nx_2^3+2\zeta\dot{x}_2=F_2\cos{\omega_ft}
\end{aligned}
\end{equation}
where $k_c$ represents the strength of coupling, $k_N$ the strength of the cubic nonlinearity, and $\zeta$ is the damping ratio. Throughout this work, we consider moderate damping, $\zeta=0.05$. We use a system with hardening nonlinearity to present our results, $k_N=1$. We consider strong coupling between the units, $k_c=5$, to avoid the overlapping of the two modes of the system. 

To investigate nonreciprocity, we analyze the steady-state response of the system under two different configurations for the input-output locations. Specifically, we define: (i) the {\it forward} configuration, where $F_1=P$ and $F_2=0$, and the output is the displacement of the right mass, $x_2^F$; and (ii) the {\it backward} configuration, where $F_1=0$ and $F_2=P$, with the output being $x_1^B$. The response of the system is reciprocal if and only if $x_2^F(t)=x_1^B(t)$.

We use the following norms to quantify the response of the system for the forward ($N^F$) and backward ($N^B$) configurations:
\begin{subequations}
\label{Norms}
    \begin{align}
    N^F &= \frac{1}{T}\int_{0}^{T}{\left({x}_{2}^F\left(t\right)\right)^2dt}\\
    N^B &= \frac{1}{T}\int_{0}^{T}{\left({x}_1^B\left(t\right)\right)^2dt}
    \end{align}
\end{subequations}
where $T=2\pi/\omega_f$ is the period of excitation. These integral-based measures are proportional to the energy in the output of the system and are commonly used in the study of nonlinear nonreciprocity~\cite{blanchard_non-reciprocity_2018}. 

We use numerical continuation, as implemented in {\sc coco}~\cite{coco}, to compute the steady-state response of the system as a family of periodic orbits that satisfy a suitable boundary-value formulation~\cite{doedel_lecture_2007}. Thus, the computed solutions are not necessarily harmonic. The stability of the response is determined by the Floquet multipliers associated with each periodic orbit. 

We define the phase of the response based on the first harmonic component of the output; {\it i.e.} the Fourier coefficients corresponding to $2\pi/\omega_f$. Appendix B provides more details on this process. This choice is motivated by the fact that we primarily operate the system in the weakly nonlinear regime where contributions from the higher-order harmonics are not significant. This is also known as the frequency-preserving response regime. For ease of reference, parameters $\phi^F$ and $\phi^B$ denote the phase of the forward and backward output displacements, respectively.

\section{Controlling the Transmitted Phase and Energy}
\label{sec:control}

We are looking for a systematic computational procedure to find parameters at which the system exhibits phase-preserving nonreciprocity ($N^F\ne N^B$, $\phi^F=\phi^B$). We achieve this by first obtaining a response characterized by phase nonreciprocity ($N^F= N^B$, $\phi^F\ne\phi^B$) and then a reciprocal response ($N^F=N^B$, $\phi^F=\phi^B$). 

\subsection{Phase Nonreciprocity}
\label{results}

\begin{figure}[h!]
\centering
\begin{subfigure}{1\textwidth}
  \centering
  \includegraphics[width=\textwidth]{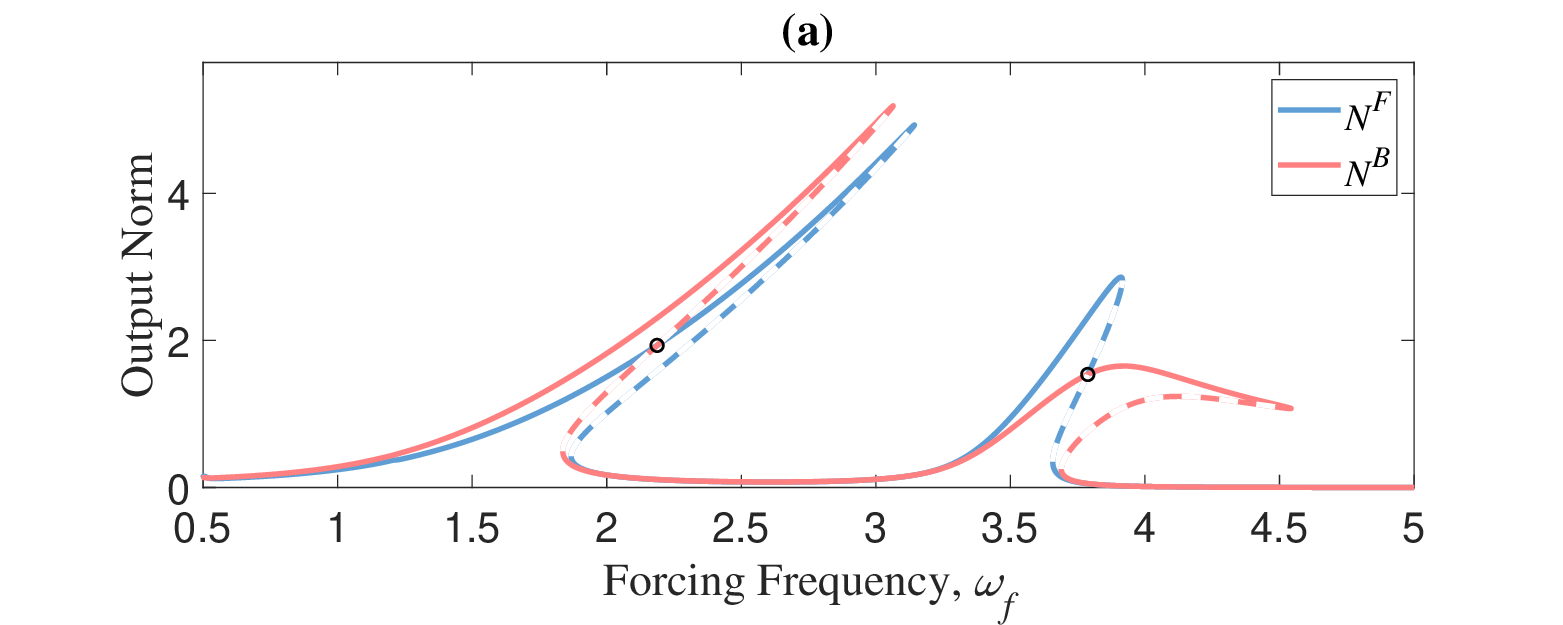}
\end{subfigure}
\vfill
\begin{subfigure}{0.45\textwidth}
  \centering
  \includegraphics[width=\textwidth]{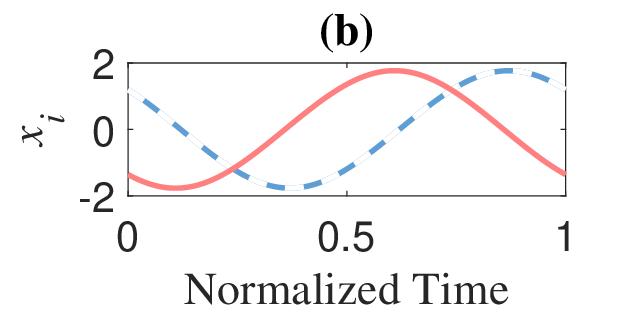}
\end{subfigure}
\hspace{0.02\textwidth} 
\begin{subfigure}{0.45\textwidth}
  \centering
  \includegraphics[width=\textwidth]{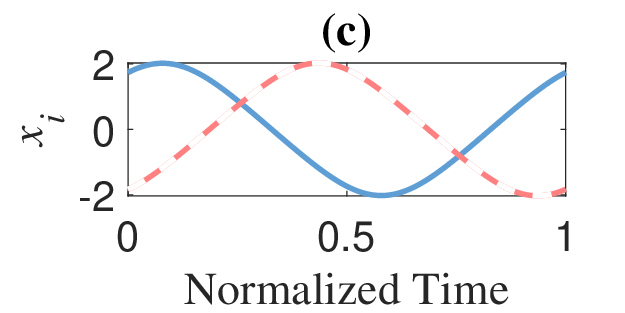}
\end{subfigure}
\caption{(a) Frequency response curves of the system with $P=2$, $\alpha=1$, $\mu=1$, and  $r=2.5$. Time response over one forcing period at the intersections points marked by black circles at (b)~$\omega_f=3.78$, and (c)~$\omega_f=2.18$. The dashed lines indicate unstable regions in the response.}
\label{fig:2}      
\end{figure}

Fig.~\ref{fig:2}(a) shows the frequency response curve of the system at $P=2$. We have used $r=2.5$ to break the mirror symmetry of the system and enable nonreciprocity in this section, while keeping $\mu=\alpha=1$.  Nonreciprocity is most conspicuous near the primary resonances because the amplitude of motion is relatively higher there. The response away from resonances is similar to that of a linear system owing to the small amplitudes, and reciprocal as a result. 

There are forcing frequencies at which the two frequency response curves intersect ($N^F=N^B$), indicating equal amplitudes in the forward and backward configurations. Figs.~\ref{fig:2}(b) and (c) show the time-domain response at the two intersection points near $\omega_f = 2.18$ and $\omega_f = 3.78$, respectively. Despite having equal amplitudes, the response at these forcing frequencies are nonreciprocal because the transmitted phase in the forward and backward configurations are different, $\phi^F\ne\phi^B$. We refer to this as the state of {\it phase nonreciprocity} and to $\Delta\phi=\phi^F-\phi^B$ as the {\it nonreciprocal phase shift} of the response. 

We note that at both intersection points identified in Fig.~\ref{fig:2} the response of the system is unstable in one of the configurations. In our approach, phase nonreciprocity is an intermediate state in finding parameters that lead to phase-preserving nonreciprocity. We can tolerate an unstable response at this stage as long as the final state of phase-preserving nonreciprocity is stable. A detailed discussion of stable states of phase nonreciprocity near the primary resonances of the system is available elsewhere~\cite{yousefzadeh_computation_2022, kogani_nonreciprocal_2024}.

\subsection{Restoring Reciprocity}

In a system that exhibits phase nonreciprocity ($N^F=N^B$), if the nonreciprocal phase shift becomes zero ($\Delta\phi=0$), then the output displacements become identical ($x_2^F=x_1^B$) and we retrieve a reciprocal response. To achieve this, a second symmetry-breaking parameter (other than $r$) is required to counterbalance the effect of the existing asymmetry and restore reciprocity~\cite{giraldo_restoring_2023}. The two symmetry-breaking parameters thus act together to maintain reciprocity in a system with broken mirror symmetry. We use the nonlinear stiffness ratio, $\alpha$, as the second symmetry-breaking parameter in this section.

\begin{figure}
\begin{subfigure}[b]{.45\textwidth}
  \includegraphics[width=\textwidth]{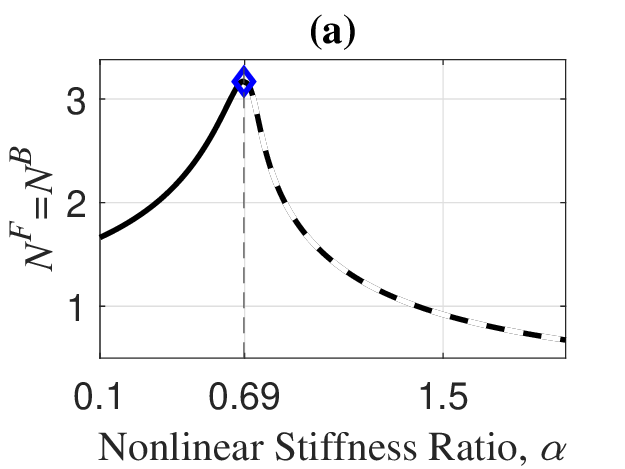}
\end{subfigure}
\hfill
\begin{subfigure}[b]{.45\textwidth}
      \includegraphics[width=\textwidth]{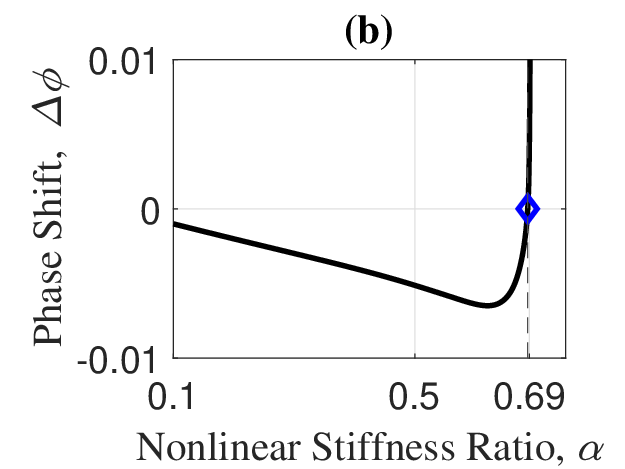}
\end{subfigure}
\hfill
\begin{subfigure}[b]{.45\textwidth}
      \includegraphics[width=\textwidth]{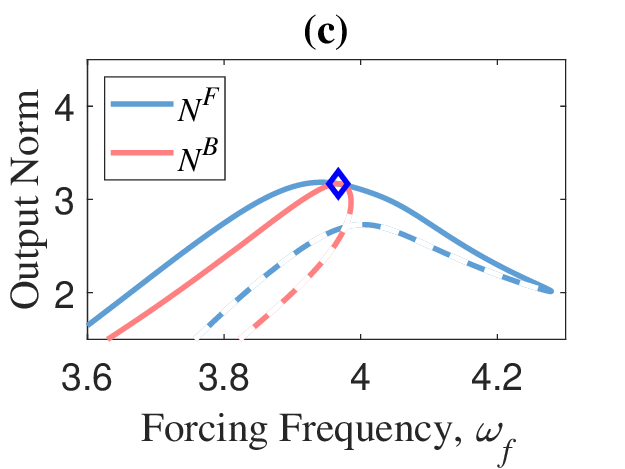}
\end{subfigure}
\hfill
\begin{subfigure}[b]{.45\textwidth}
      \includegraphics[width=\textwidth]{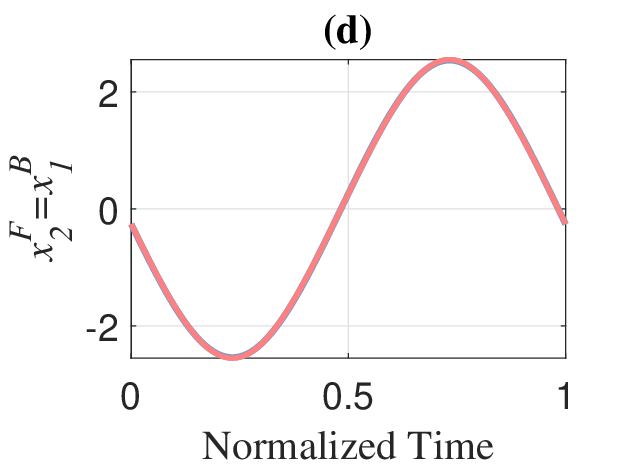}
\end{subfigure}
    \caption{(a)~Locus of phase nonreciprocity ($N^F=N^B$) as a function of the nonlinear stiffness ratio, $\alpha$. (b)~Nonreciprocal phase shift between the forward and backward configurations, $\Delta\phi=\phi^F-\phi^B$. (c)~Frequency response curves of the system for $P=2$, $r=2.5$ and $\alpha=0.69$. (d)~Time response over one forcing period corresponding to the blue diamond marker.}
    \label{fig:3}
\end{figure}

Fig.~\ref{fig:3}(a) shows the locus of phase nonreciprocity ($N^F=N^B$) as a function of $\alpha$ for the intersection point at $\omega_f=3.78$. Panel~(b) shows the variation of the nonreciprocal phase shift, $\Delta\phi$, along this locus. The blue diamond at $\alpha\approx0.69$ marks the point at which reciprocity is restored: $N^F=N^B$ and $\Delta\phi=0$.  Panel~(c) shows the frequency response curves of the system for $r=2.5$ and $\alpha=0.69$. The response of the system is nonreciprocal everywhere in this frequency range except at the point marked by the blue diamond, where the response is reciprocal, as shown in panel~(d). The reciprocal response is stable and occurs near a primary resonance of the system. More details about restoring reciprocity in a nonlinear system with broken mirror symmetry can be found elsewhere~\cite{giraldo_restoring_2023, kogani_nonreciprocal_2024}. 

We will use the reciprocal response at the diamond marker to find a family of parameters that lead to phase-preserving nonreciprocity.

\section{Phase-Preserving Nonreciprocity}
\label{sec:main}

\subsection{Influence of the forcing amplitude}

\begin{figure}
\begin{subfigure}[b]{.6\textwidth}
  \includegraphics[width=\textwidth]{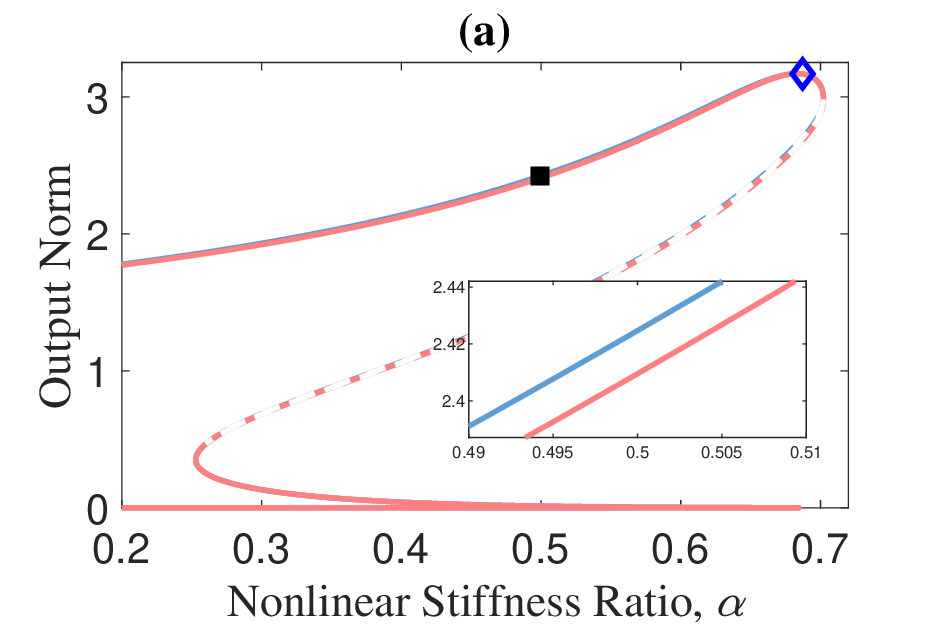}
\end{subfigure}
\hfill
\begin{subfigure}[b]{.4\textwidth}
      \includegraphics[width=\textwidth]{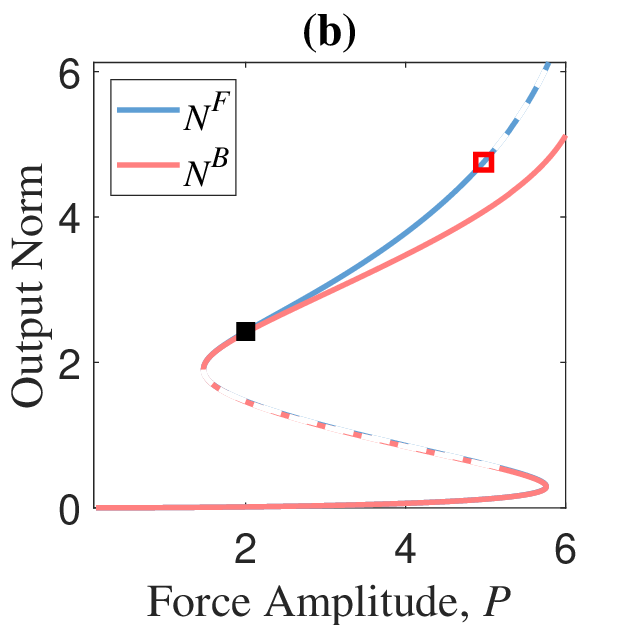}
\end{subfigure}

    \caption{Locus of phase-preserving nonreciprocity, $\Delta\phi=0$ for (a) $P=2$ and $r=2.5$ as a function of the nonlinear stiffness ratio, $\alpha$, (b) $r=2.5$ and $\alpha=0.5$ as a function of force amplitude, $P$.}
    \label{fig:4}
\end{figure}

To find a family of solutions that exhibits phase-preserving nonreciprocity, we start from the reciprocal response of a system with broken mirror symmetry (diamond marker in Fig.~\ref{fig:3}). Keeping the phase constraint, $\phi^F=\phi^B$, we compute the steady-state response manifold as a function of the nonlinear stiffness ratio, $\alpha$. This is shown in Fig.\ref{fig:4}(a). 

The response exhibits phase-preserving nonreciprocity at all the points along the locus in Fig.\ref{fig:4}(a); however, the  difference between $N^F$ and $N^B$ remains very small throughout the locus. To increase the difference in the transmitted amplitudes of the forward and backward configurations, we fix $\alpha=0.5$ (indicated by the black square marker) and increase the forcing amplitude $P$ as we keep $\Delta\phi=0$. Fig.\ref{fig:4}(b) shows that the difference between the transmitted energies can increase at higher values of the forcing amplitude. In this range of parameters, the maximum difference between $N^F$ and $N^B$ in the stable region of the response occurs near $P=5$, marked by the red square.

\begin{figure}[h!]
\centering
\begin{subfigure}{1\textwidth}
  \centering
  \includegraphics[width=\textwidth]{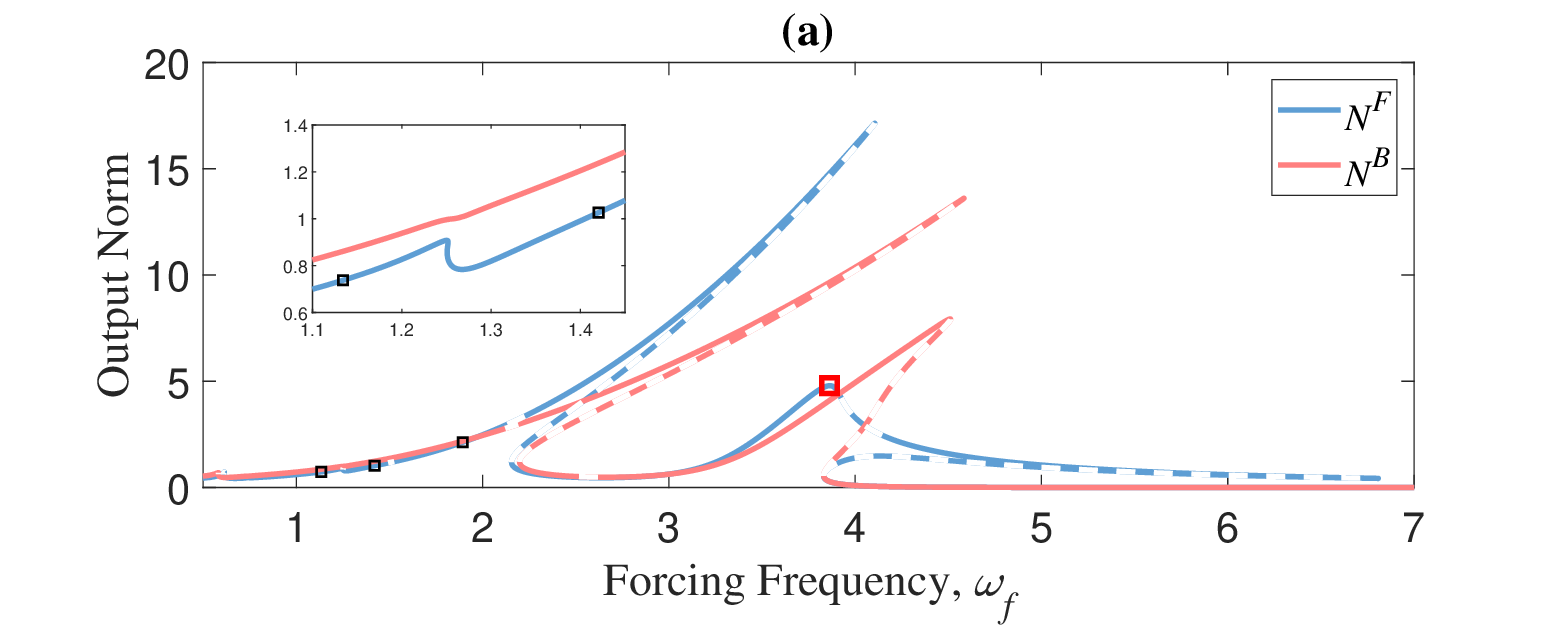}
\end{subfigure}
\vfill
\begin{subfigure}{0.23\textwidth}
  \centering
  \includegraphics[width=\textwidth]{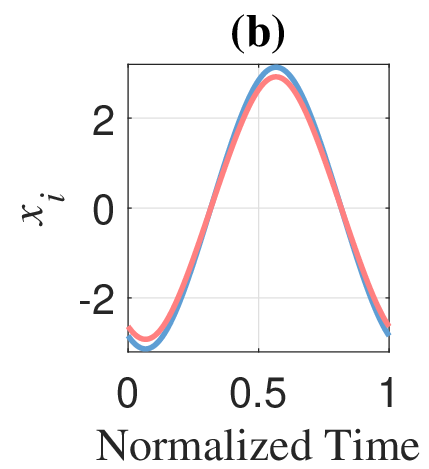}
\end{subfigure}
\hspace{0.02\textwidth} 
\begin{subfigure}{0.21\textwidth}
  \centering
  \includegraphics[width=\textwidth]{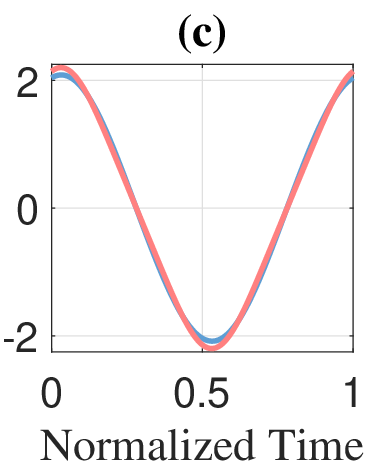}
\end{subfigure}
\hspace{0.02\textwidth} 
\begin{subfigure}{0.21\textwidth}
  \centering
  \includegraphics[width=\textwidth]{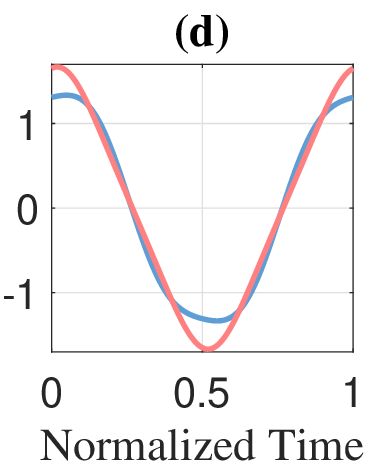}
\end{subfigure}
\hspace{0.02\textwidth} 
\begin{subfigure}{0.21\textwidth}
  \centering
  \includegraphics[width=\textwidth]{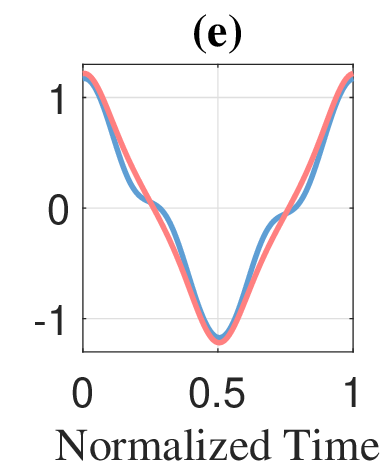}
\end{subfigure}

    \caption{(a) Frequency response curves of the system with $P=5$, $r=2.5$, and $\alpha=0.5$. Time response over one forcing period at the points indicated by square markers: (b)~$\omega_f=3.86$, (c)~$\omega_f=1.89$, (d)~$\omega_f=1.42$, and (e)~$\omega_f=1.13$.}
\label{fig:5}      
\end{figure}

Fig.~\ref{fig:5}(a) shows the frequency response curves of the system for $r=2.5$, $\alpha=0.5$, and $P=5$. The red square marks the point with phase-preserving nonreciprocity obtained in Fig.\ref{fig:4}(b). Fig.~\ref{fig:5}(b) shows the time-domain response at this point: the output displacements are harmonic and have the same phases, but the response is nonreciprocal. The amplitude of the output displacement is 7\% higher in the forward configuration in this case. 

There are three other forcing frequencies at which the system exhibits phase-preserving nonreciprocity, marked by black squares in Fig.~\ref{fig:5}(a). Panel~(c) shows the time-domain response at $\omega_f=1.89$, which is near the first primary resonance. Similar to the situation in panel~(a), the amplitude difference is small. In panels (d) and (e), phase-preserving nonreciprocity is accompanied by the appearance of higher harmonics in the forward configuration due to the proximity to a 3:1 internal resonance near $\omega_f\approx1.2$. We note that the phase difference is zero only for the first harmonics. 

\subsection{Influence of the linear stiffness ratio}

\begin{figure}[h!]
\begin{subfigure}[b]{.5\textwidth}
  \includegraphics[width=\textwidth]{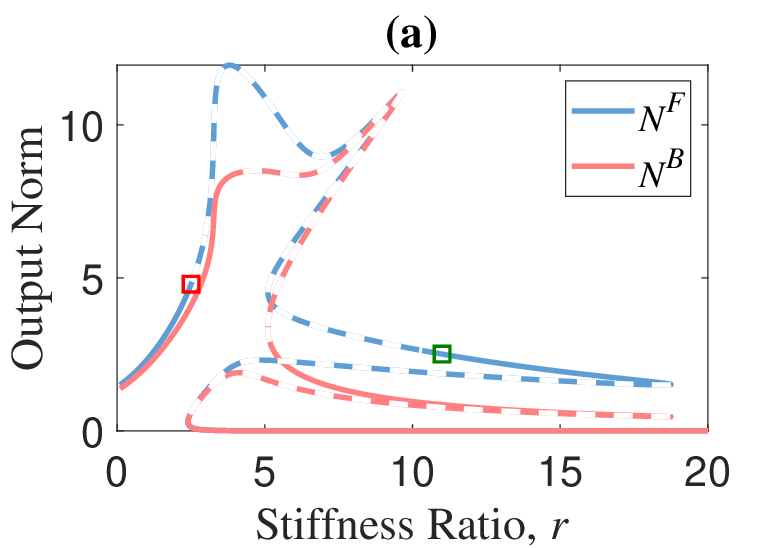}
\end{subfigure}
\hfill
\begin{subfigure}[b]{.5\textwidth}
      \includegraphics[width=\textwidth]{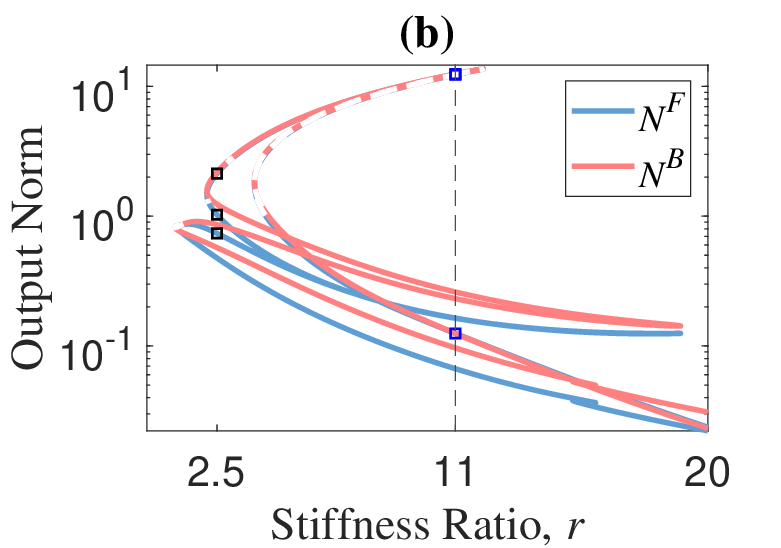}
\end{subfigure}

    \caption{Locus of phase-preserving nonreciprocity, $\Delta\phi=0$, as a function of the stiffness ratio, $r$, computed from the points indicated in Fig.~\ref{fig:5}: (a) the red marker, (b) one of the black markers.}
    \label{fig:6}
\end{figure}

We use the linear stiffness ratio, $r$, to find parameters at which the phase-preserving nonreciprocity is accompanied by a larger difference in the transmitted amplitudes. Fig.~\ref{fig:6}(a) shows the locus of phase-preserving nonreciprocity, $\Delta\phi=0$, as a function of $r$. The red square marker corresponds to the same point as in Fig.~\ref{fig:5}(a). The green square marker at $r=11$ indicates a point within the stable range of the locus at which the difference between the amplitudes of the forward and backward configurations is the largest.
Fig.~\ref{fig:6}(b) shows the locus of $\Delta\phi=0$ for the points near the first primary resonance of the system that are marked by black squares in Fig.~\ref{fig:5}(a). The locus of phase-preserving nonreciprocity forms a closed loop that passes through these points. 

\begin{figure}[h!]
\centering
\begin{subfigure}{1\textwidth}
  \centering
  \includegraphics[width=\textwidth]{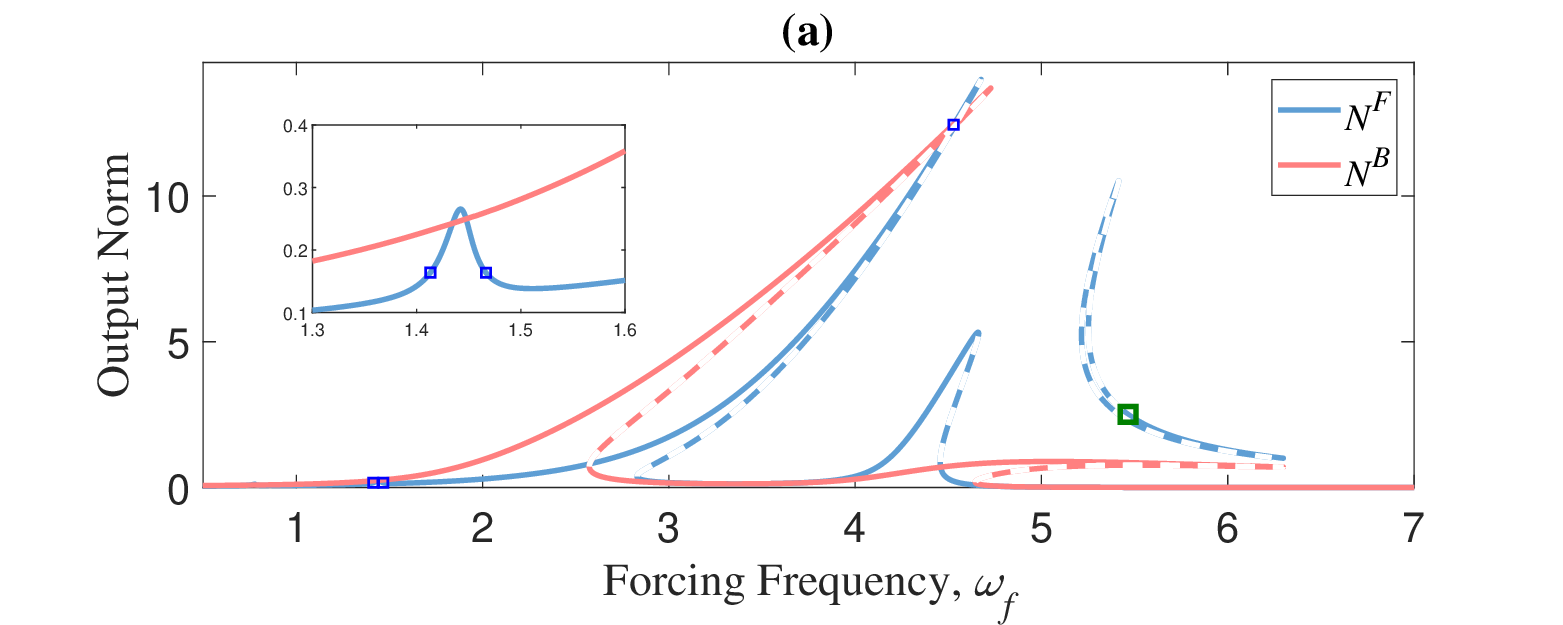}
\end{subfigure}
\vfill
\begin{subfigure}{0.23\textwidth}
  \centering
  \includegraphics[width=\textwidth]{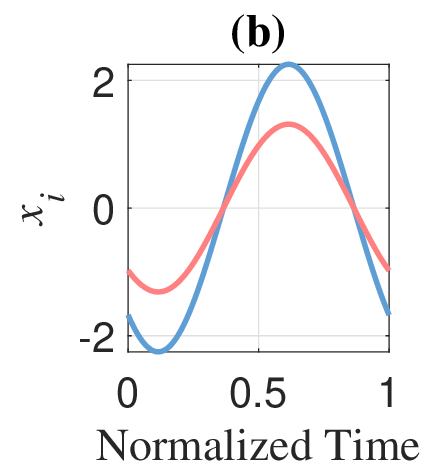}
\end{subfigure}
\hspace{0.02\textwidth} 
\begin{subfigure}{0.21\textwidth}
  \centering
  \includegraphics[width=\textwidth]{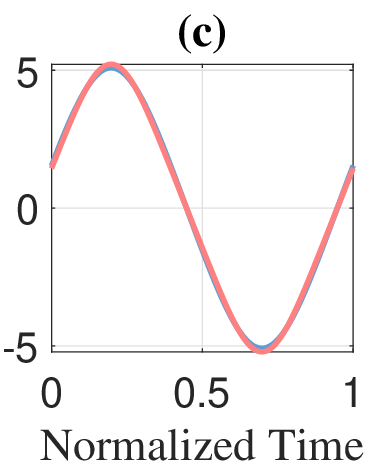}
\end{subfigure}
\hspace{0.02\textwidth} 
\begin{subfigure}{0.21\textwidth}
  \centering
  \includegraphics[width=\textwidth]{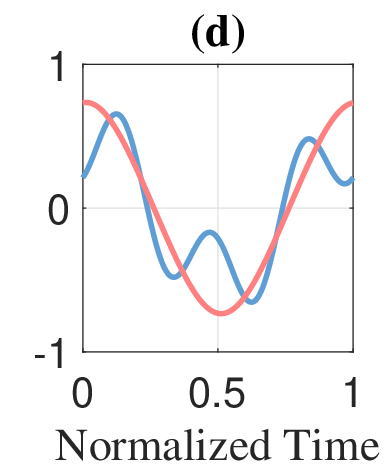}
\end{subfigure}
\hspace{0.02\textwidth} 
    \caption{(a) Frequency response curves of the system with $P=5$, $r=11$, $\mu=1, \alpha=0.5$. Time response at the points indicated by square markers: (b)~$\omega_f=5.46$, (c)~$\omega_f=4.52$ (d)~$\omega_f=1.46$.}
\label{fig:7}      
\end{figure}

Fig.~\ref{fig:7}(a) shows the frequency response curve of the system for $r=11$, $\alpha=0.5$, and $P=5$, which corresponds to the parameter values for the green square marker in Fig.~\ref{fig:6}(a). Remarkably, the green square marker lies on an isolated portion of the response curve (an {\it isola}) for the forward configuration; we could not find an isola in the response curve of the backward configuration in this frequency range. Panel~(b) shows the time-domain response of the system at the green square marker, showing a significant difference in the amplitudes of the forward and backward configurations. 

Fig.~\ref{fig:7}(a) includes other points that exhibit $\Delta\phi=0$, indicated by the blue square markers. Panel~(c) shows the time-domain response at $\omega_f=4.52$, where the difference in the transmitted amplitudes is negligible; the response happens to be almost reciprocal at this point. Panel (d) shows the time-domain response at $\omega_f=1.46$, which is one of the two square markers close to the 3:1 internal resonance of the system near $\omega_f\approx1.43$ -- see the inset in Fig.~\ref{fig:7}(a). The response at the other square marker ($\omega_f=1.41$) is very similar and is not shown. The contribution from the third harmonic in Fig.~\ref{fig:7}(d) is stronger than in Figs.~\ref{fig:5}(d) and (e). It is clear from the asymmetry of the response in Fig.~\ref{fig:7}(d) that the third harmonics in this case have different phases between the forward and backward configurations. This is because we have only enforced phase preservation for the first harmonic. 

\begin{figure}[h!]
\centering
\begin{subfigure}[b]{.7\textwidth}
      \includegraphics[width=\textwidth]{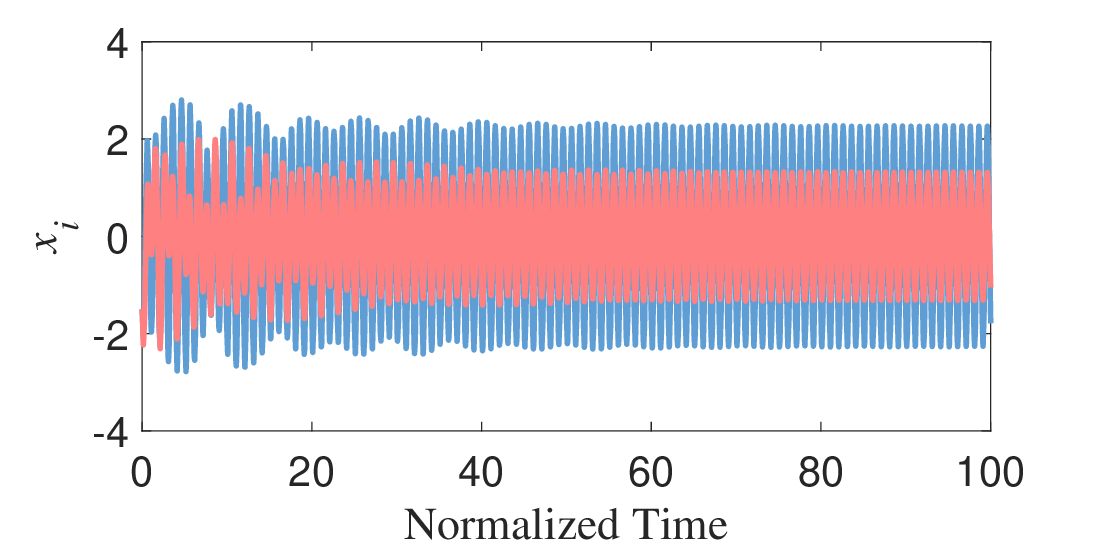}
\end{subfigure}

\caption{Transient response of the forward and backward configurations over 100 forcing periods at $\omega_f=5.46$ starting from the same initial conditions.}
    \label{fig:8}
\end{figure}

Fig.~\ref{fig:8} shows the transient response of the forward and backward configurations for initial conditions within the basin of attraction of the green blue marker in Fig.~\ref{fig:6}(a). As expected, the steady-state response in the forward configuration settles on the isola ($N^F\approx2.50$ and an amplitude of $2.25$), while the steady-state response in the backward configuration lies on the main frequency curve ($N^B\approx0.85$ and an amplitude of $1.30$).

\section{Conclusions}
\label{sec:conclusion}

We conducted a computational analysis of nonreciprocal vibration transmission in a nonlinear mechanical system with two degrees of freedom. We focused on the steady-state response of the system to external harmonic excitation. Nonlinearity appeared in the grounding nonlinear elasticity of each degree of freedom. The mirror symmetry of the system was controlled by two independent symmetry-breaking parameters. Within this context, our primary focus was on highlighting the role of transmitted phase in breaking reciprocity.

Nonreciprocity is most commonly associated with and identified by a large energy bias in the transmitted energy when the locations of the source and receiver are interchanged ($N^F\ne N^B$). This energy bias is almost always accompanied by a difference in the transmitted phase ($\phi^F\ne\phi^B$). It is possible to have response regimes in which breaking of reciprocity is solely due to a difference in the transmitted phases and not the transmitted energies ($N^F=N^B,\phi^F\ne\phi^B$). In this work, we showed that it is possible to have response regimes in which an energy bias is not accompanied by a phase bias ($N^F\ne N^B, \phi^F=\phi^B$). We provide a systematic approach for realizing such regimes of phase-preserving nonreciprocity by tuning two independent symmetry-breaking parameters of the system. 

Our findings indicate that breaking of reciprocity is most commonly accompanied by a simultaneous bias in the transmitted energy and phase. Energy bias alone, with no contribution from phase, can still lead to nonlinear nonreciprocity, albeit at very finely tuned sets of system parameters. This highlights the significant role of phase in breaking reciprocity, a feature that is often overlooked. We hope that our findings prove useful in the design and application of nonreciprocal devices in energy harvesting and mechanical signal processing. 

\section*{Acknowledgments}
We acknowledge financial support from the Natural Sciences and Engineering Research Council of Canada through the Discovery Grant program. A.K. acknowledges additional support from Concordia University. 

\nocite{*}

\bibliographystyle{unsrt} 
\bibliography{mypaper.bib}


\appendix

\section*{Appendix A: Non-dimensional Equations of Motion}
\label{appA}

The governing equations for the system in Fig.~\ref{fig:1} can be written as:
\begin{equation}
\begin{aligned}
\label{EOMraw}
M_1\ddot{x}_{1}+k_3(x_1-x_2)+k_{g1}x_{1}+k_{n1}x_1^3+c\dot{x}_{1}=f_{1}\cos{\omega_ft} \\
M_2\ddot{x}_{2}+k_3(x_{2}-x_{1})+k_{g2}x_{2}+k_{n2}x_{2}^3+c\dot{x}_{2}=f_{2}\cos{\omega_ft}
\end{aligned}
\end{equation}
where $k_3$ is the coupling stiffness. $k_{n1}$ and $k_{n2}$ are the coefficients of the nonlinear grounding stiffness for $M_1$ and $M_2$. $k_{g1}$ and $k_{g2}$ are the coefficients of the linear grounding springs for $M_1$ and $M_2$, and $c$ is the linear viscous damping connecting each mass to the ground. We divide the equations by $k_{g1}$ and introduce the non-dimensional parameters $\tau=\omega_0t$, $\omega_0^2=k_{g1}/M_1$, $\Omega=\omega_f/\omega_0$ to obtain 
\begin{equation}
\begin{aligned}
M_1\omega_0^2/k_{g1}x''_{1}+k_3/k_{g1}(x_{1}-x_{2})+x_{2i-1}+k_{n1}/k_{g1}x_{1}^3+2\zeta_gx'_{1}=f_{1}/k_{g1}\cos{\Omega\tau} \\
M_2\omega_0^2/k_{g1}x''_{2}+k_3/k_{g1}(x_{2}-x_{1})+k_{g2}/k_{g1}x_{2}+k_{n2}/k_{g1}x_{2}^3+2\zeta_gx'_{2}=f_{2}/k_{g1}\cos{\Omega\tau}
\end{aligned}
\end{equation}
where $x'=dx/d\tau=(dx/dt)/\omega_0$, $x''=d^2x/d\tau^2=(d^2x/dt^2)/\omega_0^2$, and $\zeta_g=(c\omega_0)/(2k_{g1})$. 
We define the non-dimensional displacement and force as $\bar{x}=x/d$ and $F=f/(dk_{g1})$, where $d$ is a characteristic displacement of the system. This results in 
\begin{equation}
\label{eqNonD}
\begin{aligned}
\bar{x}''_{1}+k_c(\bar{x}_{1}-\bar{x}_{2})+\bar{x}_{1}+k_{N}\bar{x}_{1}^3+2\zeta_g\bar{x}'_{1}=F_{1}\cos{\Omega\tau} \\
\mu\bar{x}''_{2}+k_c(\bar{x}_{2}-\bar{x}_{1})+r_g\bar{x}_{2}+\alpha k_{N}\bar{x}_{2}^3+2\zeta_g\bar{x}'_{2}=F_{2}\cos{\Omega\tau}
\end{aligned}
\end{equation}
where $\mu=M_2/M_1$, $k_{N}=d^2k_{n1}/k_{g1}$,  $\alpha=k_{n2}/k_{n1}$, $k_c=k_3/k_{g1}$, and $r_g=k_{g2}/k_{g1}$. 
Eq.~(\ref{eqNonD}) is the non-dimensional form of Eq.~(\ref{EOMraw}). Eq.~(\ref{eqNonD}) is the same as Eq.~(\ref{govern}) in the main text, where we have dropped the overbar in $\bar{x}$ and replaced $\tau$ with $t$, and $\Omega$ with $\omega_f$  for ease of reference.

\section*{Appendix B: Extraction of the Phase of the First Harmonic}
\label{AppB}
In this appendix, we outline the procedure for extracting the phase of the first harmonic from a periodic response using Fourier series decomposition.

A general periodic response $x(t)$ with period $T$ can be expanded in terms of its Fourier series as  

\begin{equation}
\label{eq:FourierSeries}
x(t) = a_0 + \sum_{n=1}^{\infty} \left( a_n \cos(n\omega t) + b_n \sin(n\omega t) \right),
\end{equation}
where $\omega = \frac{2\pi}{T}$ is the fundamental frequency, and the Fourier coefficients are given by  

\begin{equation}
\label{eq:FourierCoeffs}
\begin{aligned}
a_n &= \frac{2}{T} \int_0^T x(t) \cos(n\omega t) \, dt, \\
b_n &= \frac{2}{T} \int_0^T x(t) \sin(n\omega t) \, dt.
\end{aligned}
\end{equation}

To extract the phase of the first harmonic ($n=1$), we consider the first-order terms:

\begin{equation}
\label{eq:FirstHarmonic}
x_1(t) = a_1 \cos(\omega t) + b_1 \sin(\omega t).
\end{equation}

This can be rewritten in an equivalent phase-amplitude form:

\begin{equation}
\label{eq:PhaseAmplitude}
x_1(t) = A_1 \cos(\omega t - \phi_1),
\end{equation}
where the amplitude $A_1$ and phase $\phi_1$ are determined as  

\begin{equation}
\label{eq:Amplitude}
A_1 = \sqrt{a_1^2 + b_1^2},
\end{equation}

\begin{equation}
\label{eq:Phase}
\tan(\phi_1)=b_1/a_1
\end{equation}
Thus, the phase of the first harmonic is directly obtained from the ratio of the Fourier coefficients $a_1$ and $b_1$.

\end{document}